\documentclass[conference]{IEEEtran}
\IEEEoverridecommandlockouts
\usepackage{cite}

\usepackage{amsmath,amssymb,amsfonts}
\usepackage{algorithmic}
\usepackage{graphicx}
\usepackage{textcomp}
\usepackage{xcolor}

\renewcommand{\IEEEbibitemsep}{0pt plus 0.5pt}
\makeatletter
\IEEEtriggercmd{\reset@font\normalfont\fontsize{7.9pt}{8.40pt}\selectfont}
\makeatother
\IEEEtriggeratref{1}

\graphicspath{{../pdf/}{../jpeg/}{../eps/}}

\def\BibTeX{{\rm B\kern-.05em{\sc i\kern-.025em b}\kern-.08em
  T\kern-.1667em\lower.7ex\hbox{E}\kern-.125emX}}
\begin{document}
\bstctlcite{IEEE_INFOCOM:BSTcontrol}

\title{Real-Time Experimental Demonstration of Multi-band CAP Modulation in a\\
VLC System with Off-the-Shelf LEDs\\
\thanks{This work is supported by EPSRC MARVEL (EP/P006280/1).}
}

\author{Paul Anthony Haigh and Izzat Darwazeh\\
\IEEEauthorblockA{\textit{Communications and Information Systems Group, University College London, Gower Street, WC1E~6BT, United Kingdom}\\
\{p.haigh;i.darwazeh\}@ucl.ac.uk}
}

\maketitle

\begin{abstract}
We demonstrate, for the first time, $m$-CAP modulation using off-the-shelf LEDs in a VLC in real time experimental setup using field programmable gate arrays based in universal software radio peripherals (USRPs). We demonstrate transmission speeds up to $\sim$30~Mb/s can be achieved, which supports high definition television streaming.
\end{abstract}

\begin{IEEEkeywords}
Digital signal processing, field programmable gate array, modulation, real time, visible light communications
\end{IEEEkeywords}

\section{Introduction}
Advanced modulation formats are commonly used to increase transmission speeds in visible light communications (VLC), where modulation bandwidths are typically limited to several MHz \cite{wu,haigh,wang,li}. They allow, in principle, additional spectral efficiency and hence, higher data rates \cite{wang}. One of the most popular formats is carrier-less amplitude and phase (CAP) modulation, which, in 2013 \cite{wu}, was shown to outperform experimentally orthogonal frequency division multiplexing (OFDM) in terms of bit-error rate (BER). Since the report in \cite{wu}, several variations of CAP have been proposed, the most significant of which is the multi-band system ($m$-CAP) that protects against chromatic dispersion in fibres \cite{monroy}. Subsequently, $m$-CAP was adopted in VLC \cite{haigh}, it was shown that reducing subcarrier bandwidths led to tolerance of out-of-band attenuation caused by the low LED modulation bandwidths, and thus, higher spectral efficiencies.

In recent years, there have been only a limited number of studies of real-time VLC systems, where the two most recent examples are reported in \cite{li2,wang2}. Both reports utilise OFDM modulation rather than CAP due to the ease of deploying an (inverse) fast Fourier transform ((I)FFT) pair. The computational complexity of the (I)FFT pair increases with $2\log_2(N)$, where $N$ is the number of subcarriers, while $m$-CAP requires finite impulse response (FIR) pulse-shaping filters whose complexity increases with $4Lm$, where $L$ is the number of taps and $m$ is the number of subcarriers \cite{wei}. Clearly, $m$-CAP has a higher computational requirement than OFDM, however recent reports have shown that look-up tables at the transmitter can reduce this by 50\%, and when used together with new receiver architectures and pulse shapes, the computational load can be reduced by a up to $\sim$90\% \cite{haigh2}, meaning straightforward implementation.

Here, we experimentally demonstrate a real-time $m$-CAP system for the first time over a VLC system using white LEDs. Two field programmable gate arrays (FPGAs) are used in this real-time demonstration to ensure fair synchronisation and BER measurement across two independent clocks. We show that transmission speeds up to $\sim$30~Mb/s can be supported with a 20\% forward error correction (FEC) code, and $\sim$20~Mb/s with a 7\% FEC code. These rates are sufficient to support full high-definition television (HDTV) transfer, as will be demonstrated, across real VLC links using off-the-shelf LEDs.

\section{Experimental Real Time Demonstrator Setup}
\begin{figure}[h]
  \vspace{-1em}
  \centering
  \includegraphics[width=0.4825\textwidth]{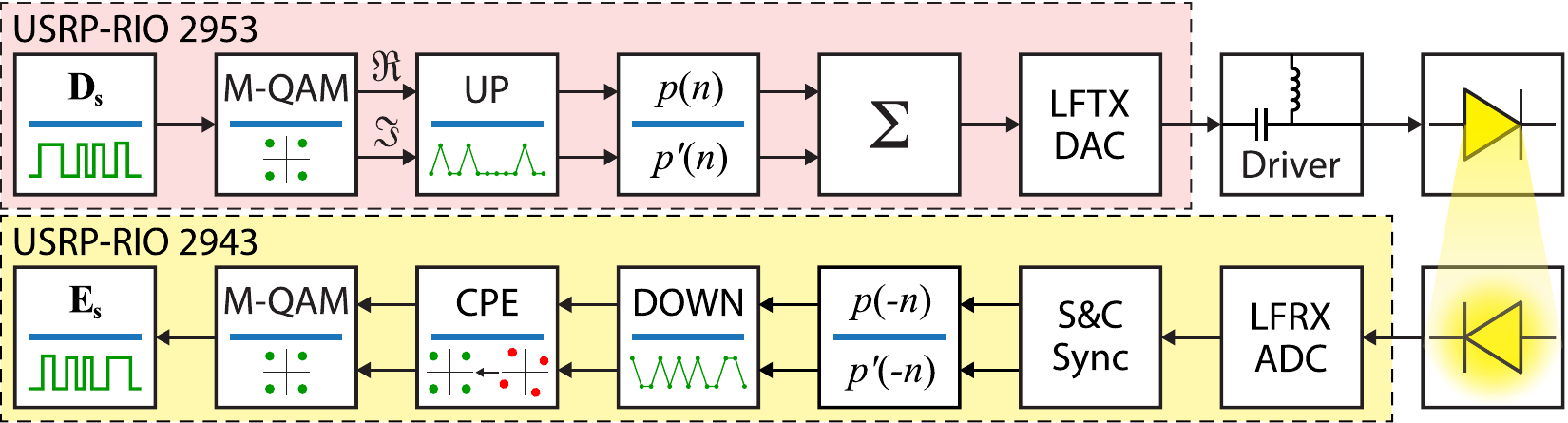}
  \caption{System block diagram (simplified)}
  \vspace{-1em}
  \label{fig:figure1}
  \vspace{-0.5em}
\end{figure}

The system block diagram is shown in Fig.~\ref{fig:figure1}. Details of the generation of $m$-CAP signals may be referred to in the literature, due to space considerations, in \cite{haigh,monroy}. We tested $m$-CAP signals, varying from $m=1$ to $m=10$, using the 4-, 16- and 64-QAM constellations for each subcarrier, with a fixed bandwidth of 6.5~MHz and roll-off factor of 0.15 as used in \cite{haigh}. The incoming information is modulated into the $m$-CAP signal format inside the transmitter FPGA, which is a Xilinx Kintex 7 (410T) contained within a National Instruments universal software radio peripheral reconfigurable input/output (USRP-RIO) 2953. The FPGA outputs the signal to an Ettus Research LFTX digital-to-analogue converter (DAC) daughtercard, which has an analogue bandwidth up to~30 MHz. The signal is then passed through a Texas Instruments THS3202 amplifier with 4 times gain, before superimposing on a 250~mA bias current.

\begin{figure}[th]
  \centering
  \includegraphics[width=0.4825\textwidth]{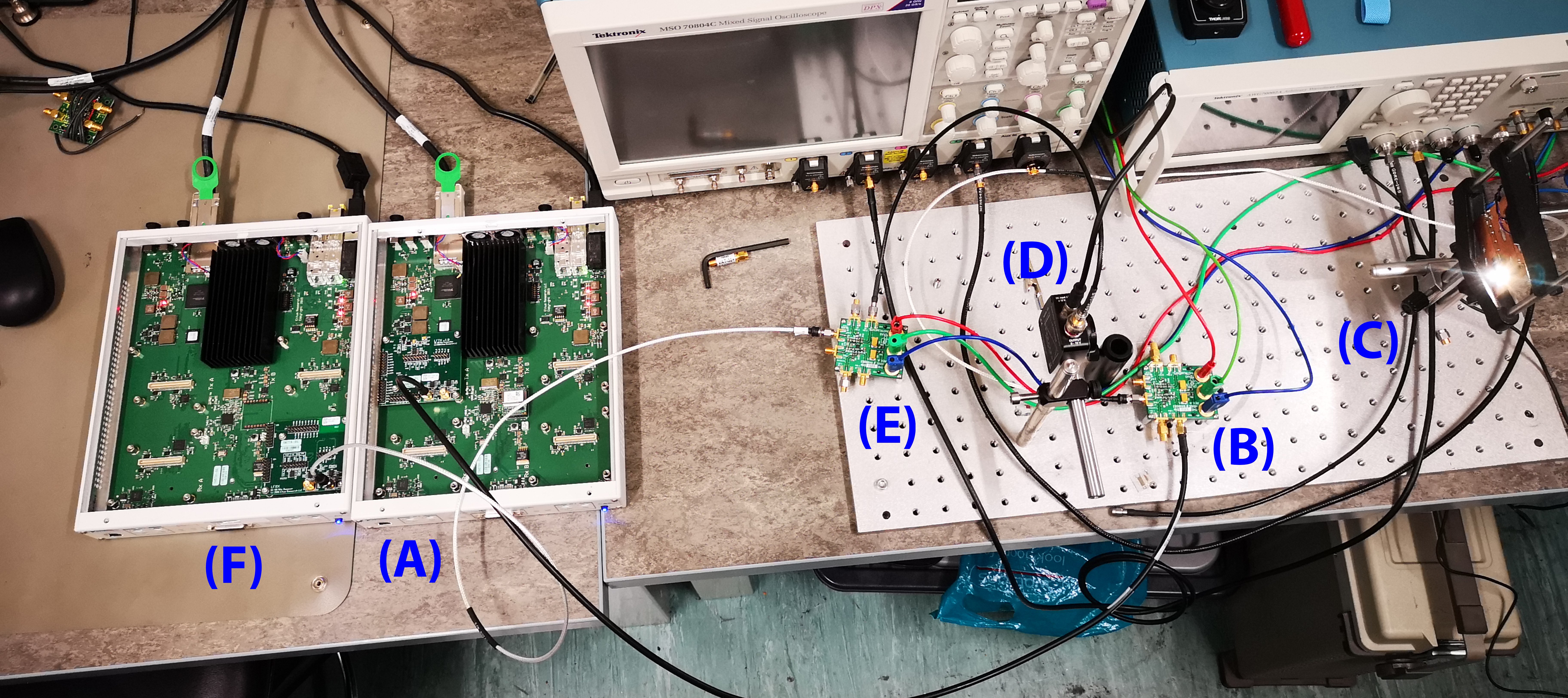}
  \caption{Photograph of demonstrator test setup: (A) transmit USRP-RIO; (B) transmitter amplifier; (C) transmit driver and LED; (D) photodetector; (E) receiver amplifier and (F) receiver FPGA}
  \vspace{-1em}
  \label{fig:figure2}
\end{figure}

The biased signal that intensity modulates the Osram Golden Dragon LED (optoelectronic characteristics can be referred to in \cite{haigh3}) for transmission over a 0.5~m link, which is fixed to maintain a compact demonstrator, however operation can also be supported at 1~m distances. No bulk optics, such as lenses, are used and a Thorlabs PDA100A2 packaged receiver is used, which consists of a silicon $p$-$i$-$n$ photodiode and an in-built transimpedance amplifier. The amplifier output is further amplified ($10\times$) by a Texas Instruments THS3202 amplifier to maximize symmetrical swing at the receiver input.
The receiver is a National Instruments USRP-RIO 2943 with the same Xilinx Kintex 7 FPGA on-board, but running independently from the transmitter. An Ettus Research LFRX analogue-to-digital converter (ADC) daughtercard (0-30~MHz) is used to digitise the data and direct it to the FPGA, for demodulation and BER measurements. Synchronisation is obtained using the Schmidl and Cox \cite{schmidl} method while the demodulation of $m$-CAP signals may be referred to in \cite{haigh,monroy}. A photograph of the experimental setup is shown in Fig.~\ref{fig:figure2}.

\section{Results}
\begin{figure}[th]
  \centering
  \includegraphics[width=0.45\textwidth]{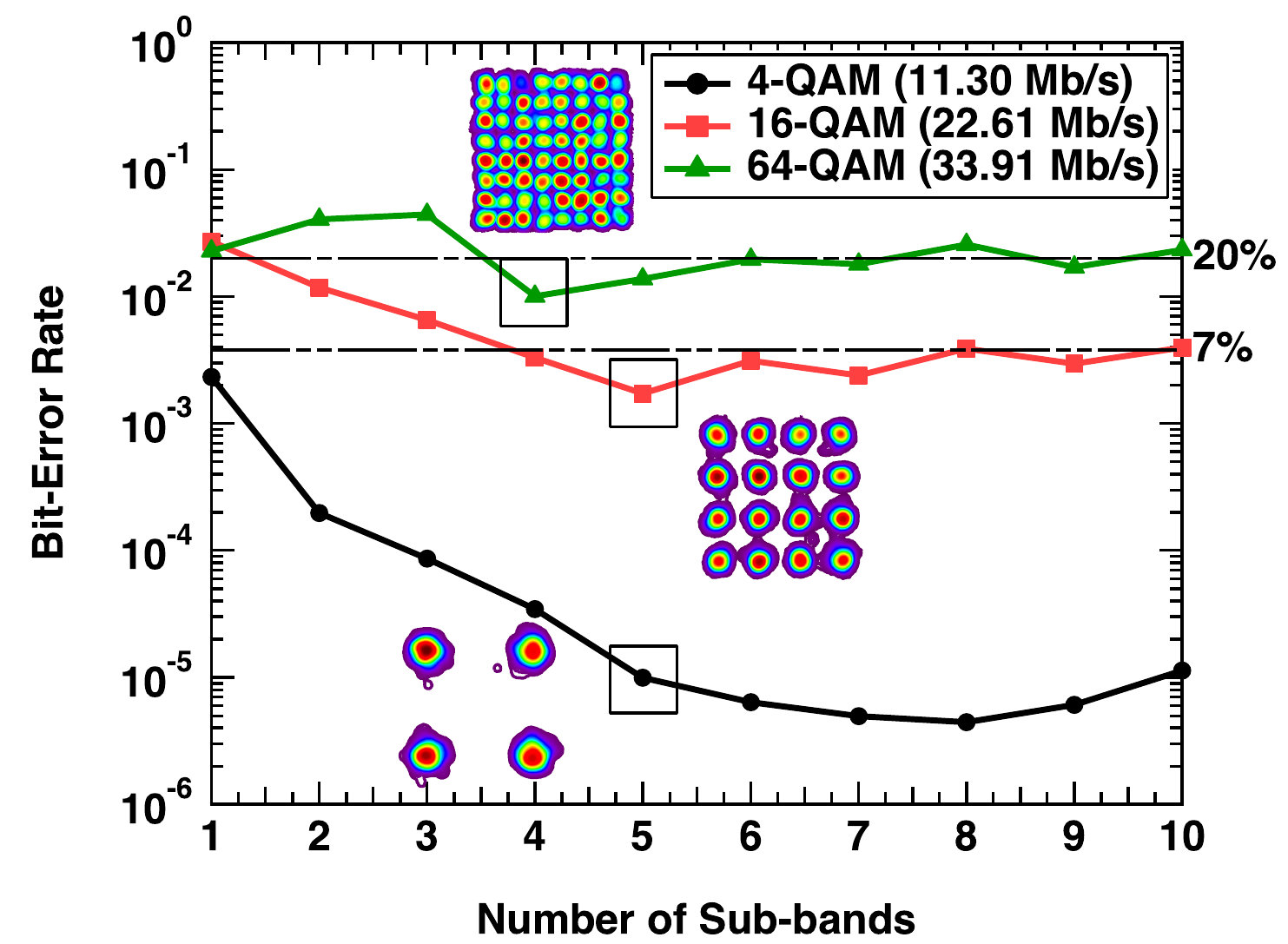}
  \vspace{-1em}
  \caption{Measured real time BER performance of 4-. 16- and 64-QAM. Transmission can be supported with 4- and 16-QAM with a 7\% FEC code at 11.3~Mb/s and 22.61~Mb/s, respectively, and with a 20\% FEC code for 16-QAM at 33.91~Mb/s.}
  \vspace{-1em}
  \label{fig:figure3}
\end{figure}

\begin{figure}[th]
  \centering
  \includegraphics[width=0.45\textwidth]{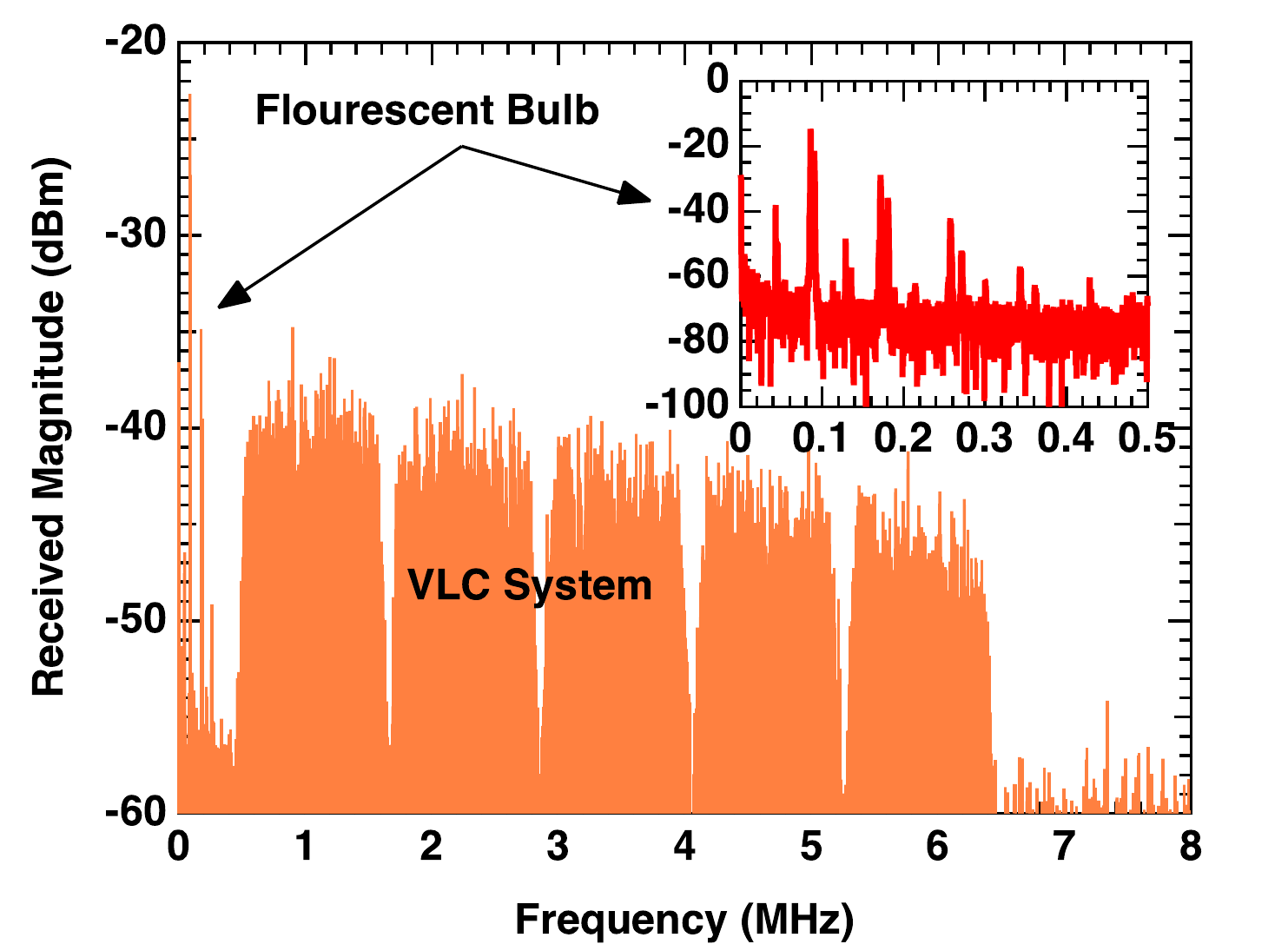}
  \vspace{-1em}
  \caption{The received electrical signal spectrum for a 5-CAP system, and inset, the flourescent component (0-500~kHz).}
  \vspace{-1em}
  \label{fig:figure4}
\end{figure}

The total accumulated BER is shown in Fig.~\ref{fig:figure3}, and $>10^6$ symbols were transmitted. A clear trend is evident; a low order of $m$ indicates a high BER, which reduces as $m$ increases. The reason for this is due to inter-symbol interference (ISI) caused by attenuation of high frequencies as documented in \cite{haigh}. However, there is another contributor to this, which is the slight impact of fluorescent light bulbs present in the laboratory. A dark environment is unrealistic in a real-world scenario, and hence, a 500~kHz frequency offset was introduced to the signals allow co-existence of fluorescent bulbs alongside the VLC system, as illustrated in Fig.~\ref{fig:figure4}, which resulted in minor intermodulation. This effect reduces with increasing m as the aggregated roll-off at the bandwidth edges becomes sharper. The transmission rate is sufficient to demonstrate HDTV streaming.

\section{About the Demonstrator and Conclusion}
This demonstration is a self-contained and flexible test environment that supports HDTV video streaming in real time. Modulation formats can be adapted in real-time, albeit with an interruption in service before data recovery. Attendees will be able to block physically the optical path and observe the impact of a degradation in BER on the video quality. Wireless internet access will also be available via the VLC system through the demonstrator, which reports a real-time $m$-CAP VLC system, for the first time.

\bibliographystyle{IEEEtran}
\bibliography{IEEE_INFOCOM}
\bstctlcite{IEEE_INFOCOM}

\end{document}